\documentstyle[aps,eqsecnum,tighten,epsf]{revtex}
\newcommand{\be}{\begin{equation}}
\newcommand{\ee}{\end{equation}}
\newcommand{\ba}{\begin{eqnarray}}
\newcommand{\ea}{\end{eqnarray}}
\newcommand{\bb}{\begin{thebibliography}}
\newcommand{\eb}{\end{thebibliography}}

\begin{document}

\begin{center}
{ \Large \bf  ON THE $ \tau \to (a_1 h)^- \nu_{\tau}$ DECAYS } \\
\vspace*{0.5cm}
{\large  K.R.Nasriddinov, B.N.Kuranov, U.A.Khalikov } \\
\end{center}
{\large
\begin{center}
{\ Institute of Nuclear Physics, Academy of Sciences of Uzbekistan,\\
    pos. Ulugbek, Tashkent,702132 Uzbekistan } \\
\end{center}
\begin{center}
{\large  T.A. Merkulova  } \\
\end{center}
{\large
\begin{center}
{\ Joint Institute for Nuclear Research, Dubna, 141980 Russia }\\
\end{center}
}
\begin{quote}
   The $ \tau \to (a_1 h)^- \nu_{\tau}$ decays of the $\tau$-lepton
are studied using the method of phenomenological chiral Lagrangians.
The expression of weak hadronic currents between pseudoscalar and
axial-vector meson states is obtained. Calculated
partial widths for these decays are compared with the available
experimental data.\\
PACS number(s): 13.35.Dx, 12.39.Fe
\end{quote}

{\large
In this paper the $ \tau \to (a_1 h)^- \nu_{\tau}$ decays of the
$\tau$-lepton are studied using the method of phenomenological chiral
Lagrangians (PCL's) [1]. The main uncertainty in the study of these
decay channels is connected with weak hadron currents. Therefore such
decay channels are a unique "laboratory" for verification of weak
hadron currents between pseudoscalar and axial-vector meson states and
investigations of these decays are of interest. Note that the hadron
decays of the $\tau$-lepton up to three pseudoscalar mesons in the
final state [2,3] and also $ \tau \to VP \nu_{\tau}$ decays [4,5] have
been studied in the framework of this method.

In the PCL, the weak interaction Lagrangian, has the form
\vspace*{0.3cm}
\be
L_W =\frac{G_F}{\sqrt{2}} {J^h_{\mu} l^+_{\mu} + H.c.},
\ee \vspace*{0.3cm} where  $G_{F}\simeq10^{-5}/m^{2}_{P}$ is the Fermi
constant,\\
$ l_{\mu} = \bar {u_{l}}\gamma_{\mu}(1+\gamma_{5})u_{\nu_{l}}$ is the
lepton current, and hadron currents have the form [1]
\vspace*{0.5cm}
$$ J^h_{\mu} = J^{1+i2}_{\mu} \cos \Theta_c + J^{4-i5}_{\mu} \sin
\Theta_c,$$ where $\Theta_c$ is the Cabibbo angle.\\

Weak hadron currents between pseudoscalar and axial-vector
meson states are obtained by including the gauge fields of these
mesons in covariant derivatives [6]:
\be
\partial_{\mu} \to \partial_{\mu} + igv_{\mu}V + iga_{\mu}A,
\ee
here
$v^i_{\mu}$ and $a^i_{\mu}$ are the fields of the $1^-$ and
$1^+$ - mesons,
$V_i = \lambda_i I/2$, and $A_i = V_i \gamma_5$ are the vector
and axial-vector generators of the $SU(3)\times SU(3)$ group,
respectively. \\
In this method the hadron currents are defined as [6]
\vspace*{0.3cm}
\be
i \lambda^i J^{i}_{\mu} = F^2_{\pi}
e^{i \xi A}(\partial_{\mu} + igv_{\mu}V + iga_{\mu}A)e^{-i \xi A},
\ee
here $\xi=\frac{1}{F_{\pi}} \lambda^i \varphi^i$,
$F_{\pi}$ = 93 MeV, $\varphi^i$ represent the fields of the $0^-$
mesons, and $g$ is the "universal" coupling constant,
which is fixed from the experimental $\rho \to \pi\pi$ decay width
\vspace*{0.1cm}
$$ \frac{g^2}{4 \pi} \simeq3.2. $$

The weak hadron currents between pseudoscalar and axial-vector
meson states obtained in this way have the form
\vspace*{0.3cm}
\be
J^{i}_{\mu} = F_{\pi}ga^b_{\mu}  \varphi^cf_{bci}.
\ee

Axial-vector and vector meson currents are defined as
\vspace*{0.3cm}
\be
J^{i}_{\mu} = \frac{m^2_v}{g}{v^i_{\mu}} + \frac{m^2_a}{g}{a^i_{\mu}},
\ee
where $m_v$ and $m_a$ are the masses of vector and axial-vector mesons
, respectively.

The strong interaction Lagrangian of axial-vector mesons with vector
and pseudoscalar mesons is obtained also by this way and has the
form [2]
\vspace*{0.3cm}
\be
L_S(1^+, 1^-, 0^-) = -F_{\pi}g^2 f_{klm}a^k_{\mu}v^l_{\mu} \varphi^m.
\ee
\vspace*{0.3cm}

The decay amplitudes for these channels can be written as [7]
$$ M\left( \tau(k_{\tau}) \to a_1(p)h(p_1) \nu_{\tau}(k_{\nu})\right)=
   G_F\epsilon^{\lambda}_{\mu}\bar
   U(k_{\nu})\gamma_{\mu}[ f_1 + g_1\gamma_5 + \hat p\left( f_2 + g_2
	 \gamma_5\right)  $$
$ + \hat p_1\left( f_3 + g_3\gamma_5\right)] U(k_{\tau}), $ \\
\vspace*{0.3cm}
where $\epsilon^{\lambda}_{\mu}$ is the polarization vector of
$1^{\pm}$ mesons, $f_{i}$ and $g_{i}$ are the form
factors that depend on the final state momenta;
$q=k_{\tau}-k_{\nu}=p + p_1$, and $k_{\tau}$, $k_{\nu}$ are the lepton
four-momenta ($\hat p_{i} \equiv p_{i\mu}\gamma^{\mu}$).

Using these Lagrangians we calculated the partial widths of the
$\tau \to (a_1 h)^- \nu_{\tau}$ decays by means of the TWIST code [8].
The results are shown in the Table I. In columns I and II are
listed the results without $1^-$ contributions,
and with the vector $1^-$-meson contributions, respectively.

These decay channels get contributions from the
$\rho(770)$-, $\rho(1450)$-, and $\rho(1700)$- vector intermediate
meson states which have widths of 150, 310, and 240 MeV,
respectively. Note that the contribution of the $\rho(1450)$- and
$\rho(1700)$- mesons to
the partial widths dominate those of the $\rho(770)$ ones.

Table I shows that the result obtained for the
$\Gamma(\tau^- \to (a_1 h)^- \nu_{\tau})=0.79 \times 10^{10} sec^{-1}$
decay channels without taking into account
the vector $1^-$ contributions are in good agreement with available
experimental data [9]: \\
$\Gamma(\tau^- \to (a_1 h)^- \nu_{\tau})<6.9 \times 10^{10} sec^{-1}$.\\
Note that the calculated partial widths with taking into account
$1^-$ contributions lie above this experimental value. In these
calculations we used, as in Ref.s [2-5], the same $g$-coupling
constant, according to Eq. (2), for all the vector intermediate
meson states. Indeed, it is a rough appoximation and it was more
appreciable in study of such $\tau$ lepton rare decays than in Ref.s
[2-5]. Therefore, it would be expedient to present $g$-coupling
constant in Eq. (2) in a matrix form so that various decay channels
have their own coupling constants. Though at present we have shortage
of experimental data on the vector intermediate mesons, but there are
some theoretical attempts to determine coupling constants of such
mesons (see Ref.s [10,11]). And taking into account corresponding
coupling constants in future would allow us to describe these decays
more correctly compared to these calculations.

Note that according to Eq.(4) the partial widths
of the $ \tau^{-} \to a^-_1 \eta \nu_{\tau}$ and
$ \tau^{-} \to a^-_1 \eta' \nu_{\tau}$ decays are equal to zero
in the PCL method; as in Ref.[4], these decay channels can be realized
via effects of secondary importance [5].

Thus, the expression of weak hadronic currents between pseudoscalar
and axial-vector meson states Eq. (4) obtained by including the gauge
fields of axial-vector and vector mesons in covariant derivatives
allow us to describe the $ \tau \to (a^-_1 h)^- \nu_{\tau}$ decays
in satisfactory agreement with available experimental data.
Determination of corresponding coupling constants in Eq. (4) would
allow us to calculate these decay probabilities with high accuracy.
Probably, contributions from the vector intermediate mesons which are
very sensitive to $g$ would be in satisfactory agreement with the
experimental data.

We would like to thank F. Hussain, M. Fabbrichesi, D. V. Sao,
S. Azakov, M.M.Musakhanov, and A.Rakhimov for interest in this study
and for useful discussions. One of authors (KRN) would like to thank
Prof. S. Randjbar-Daemi for hospitality at the ICTP during the course
of this work.\\

\newpage
{
\begin{center}
{\large   Table 1. The partial widths (in $sec^{-1}$)
for the $ \tau^- \to (a_1 h)^- \nu_{\tau}$ decays.}    \\
\vspace*{0.5cm}
\begin{tabular}{|r|c|c|c|}\hline
Decays&I&II&Experiment [9]\\\hline
$ \tau^{-} \to a^0_1 \pi^- \nu_{\tau}$&
$0.39 \times 10^{10}$&$0.5 \times 10^{11}$&$----$\\
$ \tau^{-} \to a^0_1 K^- \nu_{\tau}$&
$0.73 \times 10^5$&$0.92 \times 10^4$&$----$\\
$ \tau^{-} \to a^-_1 \pi^0 \nu_{\tau}$&
$0.40 \times 10^{10}$&$0.52 \times 10^{11}$&$----$\\
$ \tau^{-} \to a^-_1 \bar K^0 \nu_{\tau}$&
$0.92 \times 10^5$&$0.19 \times 10^5$&$----$\\\hline
\end{tabular}
\end{center}
\newpage
\begin{center}
\vspace*{0.5cm}
Fig.1. Diagrams for the $ \tau^- \to (a_1 h)^- \nu_{\tau}$ decays,
       here W and S are the vertices of weak and strong interactions,
			 respectively. (a) is without the pole conrtibution of $1^{-}$
			 mesons and (b) includes these pole contributions.
\end{center}

}
\end{document}